Skyrmion Formation Induced by Antiferromagnetic-enhanced Interfacial Dzyaloshinskii Moriya Interaction

Author: Marco Chung Ting Ma[1], Yunkun Xie[2], Howard Sheng[3], S. Joseph Poon[1], and Avik Ghosh[2]

[1]Department of Physics, University of Virginia, Charlottesville, Virginia 22904 USA

[2]Department of Electrical and Computer Engineering, University of Virginia, Charlottesville, Virginia 22904 USA

[3]Department of Physics and Astronomy, George Mason University, Fairfax, Virginia 22030 USA

**Abstract**

Neél skyrmions originate from interfacial Dzyaloshinskii Moriya interaction (DMI). Recent studies have explored using ferromagnet to host Neél skyrmions for device applications. However, challenges remain to reduce the size of skyrmion to near 10 nm. Amorphous rare-earth-transitional-metal ferrimagnets are attractive alternative materials to obtain ultrasmall skyrmions at room temperature. Their intrinsic perpendicular magnetic anisotropy and tunable magnetization provides a favorable environment for skyrmion stability. In this work, we employ atomistic stochastic Landau-Liftshitz-Gilbert (LLG) algorithm to investigate skyrmions in GdFe within the interfacial DMI model. Despite the rapid decay of DMI away from the interface, small skyrmions of near 10 nm are found in thick ~ 5 nm amorphous GdFe film at 300K. We have also considered three scenarios for the sign of DMI between Gd-Fe pair. It is revealed that antiferromagnetic coupling in the ferrimagnet plays an important role in enhancing the effect of interfacial DMI and to stabilize skyrmion. These results show that ferrimagnets and antiferromagnets with intrinsic antiferromagnetic couplings are appealing materials to host small skyrmions at room temperature, which is crucial to improve density and energy efficiency in skyrmion based devices.

**Introduction**

Magnetic skyrmions are topologically protected spin textures. Their potentials in advancing memory density and efficiency have drawn extensive investigations in recent years[1-24]. In magnetic materials, skyrmions are stabilized through Dzyaloshinskii Moriya interaction (DMI)[25-26]. DMI is generated by either intrinsic or interfacial effect. Intrinsic DMI arises in non-centrosymmetric crystal, such as B20 alloys, where Bloch skyrmions are found to exist in MnSi and FeGe at low temperature[13-14]. Interfacial DMI originates from interfacial layer with strong spin-orbit coupling. Multilayer stacks, such as Ir/Fe/Co/Pt and Pt/Co/Ta, are found to host ~ 50 nm Neél skyrmions at room temperature[15-16]. Several challenges remain in developing skyrmion based memory and logic devices. For example, further reduction in skyrmion sizes is needed to optimize skyrmion based devices. However, the stability of small skyrmion at room temperature becomes a problem. Thicker magnetic layers are required to increase stability[17-18]. For ferromagnet/heavy metal multilayer stacks, increase in thickness of magnetic layer can lead to the loss of interfacial anisotropy and the reduction of the strength of DMI[52-55]. Both are critical for skyrmion formations. Moreover, skyrmions Hall effect can provide great challenges on moving skyrmions in electronics devices[19-22]. To overcome these challenges, one needs to explore more materials.

Amorphous rare-earth-transitional-metal (RE-TM) ferrimagnet is one of the potential materials to overcome these challenges. Several properties of RE-TM alloys provide favorable environment to host small skyrmions at room temperature. Their Intrinsic perpendicular magnetic anisotropy (PMA) [27-30] gives a crucial advantage in stabilizing small skyrmion by allowing the use of thicker films (~ 5 nm). However, the effectiveness of interfacial DMI decreases significantly away from the interface[52-55]. Besides PMA, the magnetization of RE-TM alloys vanishes at the compensation temperature[31]. With near zero magnetization, the skyrmion Hall effect is vastly reduced. Another advantage of RE-TM alloys is the access to ultrafast switching[32-39]. Recently, all-optical switching helicity-dependent has been demonstrated in RE-TM alloys using a circularly polarized laser[32-35]. This gives an additional tool to control spins in future devices. RE-TM alloys have begun to draw interest in the field of skyrmions research. Large skyrmions of ~ 150 nm have been observed in Pt/GdFeCo/MgO[23], and skyrmion bound pairs are found in Gd/Fe multilayers[24]. Further tuning is needed to reduce the size of skyrmion in RE-TM alloys. To guide experiments, numerical model has served as an important tool, especially for complex systems such as RE-TM alloys[34,40-44]. Several methods, such as atomistic Landau-Liftshitz-Gilbert (LLG) algorithm[34,40-43] and micromagnetic Landau-Lifshitz-Bloch (LLB) algorithm[44], has been employed to provide deeper understanding of magnetic properties in RE-TM alloys.

In this study, atomistic LLG algorithm[34,40-43] is employed to study properties of skyrmions in GdFe with interfacial DMI. Although the sign of DMI at ferromagnets/heavy metal interface is well studied[44-51], the sign of DMI involved ferrimagnet remains complex. Here, we consider three scenarios for the DMI between Gd and Fe ($D_{Gd-Fe}$). First, the influence of DMI between antiferromagnetic pair is excluded by setting it to zero ($D_{Gd-Fe}$ = 0). Second, DMI between antiferromagnetic pair is set to the same sign as DMI between ferromagnetic pair, where $D_{Gd-Fe}$ > 0. Finally, the case of $D_{Gd-Fe}$ < 0 is considered. Furthermore, to incorporate DMI being an interfacial effect, an exponential decay DMI is utilized. Simulation results find that near 10 nm skyrmions remain robust in ~ 5 nm GdFe at room temperature. This demonstrates that interfacial DMI remains prominent in thicker ferrimagnet samples, which is critical in stabilizing small skyrmions at room temperature.

**Simulation Model**

The classical atomistic Hamiltonian H in Eq. (1) is employed to investigate magnetic textures in amorphous ferrimagnets.

$$H = -\frac{1}{2}\sum_{<i,j>} J_{ij}\boldsymbol{s_i} \cdot \boldsymbol{s_j} - \frac{1}{2}\sum_{<i,j>} \boldsymbol{D_{ij}} \cdot (\boldsymbol{s_i} \times \boldsymbol{s_j}) - K_i(\boldsymbol{s_i} \cdot \widehat{\boldsymbol{K_i}})^2$$

$$-\mu_0\mu_i\boldsymbol{H_{ext}} \cdot \boldsymbol{s_i} - \mu_0\mu_i\boldsymbol{H_{demag}} \cdot \boldsymbol{s_i} \quad (1)$$

Where $\boldsymbol{s_i}, \boldsymbol{s_j}$ are the normalized spin at site i, j respectively, $\mu_i\mu_j$ are the atomic moment at site i, j respectively. Atomic moment is absorbed into the following constant, $J_{ij} = \mu_i\mu_j j_{ij}$ is the exchange interaction, $\boldsymbol{D_{ij}} = \mu_i\mu_j \boldsymbol{d_{ij}}$ is the DMI interaction and $K_i = \mu_i k_i$ is the anisotropy. $\boldsymbol{H_{ext}}$ and $\boldsymbol{H_{demag}}$ is the external field and demagnetization field respectively.

Only nearest neighbor interactions are considered in exchange and DMI interactions. Periodic boundary condition is enforced in x and y direction. To find the ground state, spins are evolved under the stochastic Landau-Lifshitz-Gilbert (LLG) Equation as shown in Eq. (2), and the constant parameters used in the simulation are listed in **Table 1**.

$$\frac{d\boldsymbol{M}}{dt} = -\frac{\gamma}{1+\alpha^2}\boldsymbol{M} \times (\boldsymbol{H}_{eff} + \boldsymbol{\xi}) - \frac{\gamma\alpha}{(1+\alpha^2)M_s}\boldsymbol{M} \times [\boldsymbol{M} \times (\boldsymbol{H}_{eff} + \boldsymbol{\xi})] \quad (2)$$

Where $\gamma$ is the gyromagnetic ratio, $\alpha$ is the Gilbert damping constant, $\boldsymbol{H}_{eff}$ is the effective field, $\boldsymbol{\xi}$ is the Gaussian white noise term for thermal fluctuation and $M_s$ is the saturation magnetization.

To incorporate the amorphous short range order, an amorphous structure of a 1.6 nm x 1.6 nm x 1.6 nm box containing 250 atoms is generated from ab initio molecular dynamics calculations by Sheng et al.[56]. **Fig. 1** shows a plot of RE and TM atoms in the amorphous structure. Replicas of this box (32 x 32 x 1) are placed next to each other to expand the simulated sample to 50.7 nm x 50.7 nm x 1.6 nm and 256000 atoms. For a 4.8 nm thick sample, replicas of the box are also placed in z-direction, and the total number of atoms is 768000.

**Results and Discussion**

With ferromagnetic DMI ($D_{Gd-Gd}$ and $D_{Fe-Fe}$) remains positive, three scenarios of antiferromagnetic DMI ($D_{Gd-Fe}$) are considered. Samples of 50.7 nm x 50.7 nm x 1.6 nm are simulated using atomistic LLG equation from Eq. (2) at 0 K. **Fig. 2** shows the equilibrium spin configurations at 0 K. For $D_{Gd-Fe} = 0$ and $D_{Gd-Fe} < 0$, skyrmion's radius increases as DMI increases and becomes stripe at large DMI value, which behaves similar to a ferromagnet[17-18]. On the other hand, with $D_{Gd-Fe} > 0$, same sign as $D_{Gd-Gd}$ and $D_{Fe-Fe}$, the trend of skyrmion sizes is somewhat different. At small DMI value, skyrmion's radius increases as DMI increases. However, at large DMI value, skyrmion's radius decreases as DMI increases, which is different from what observed in $D_{Gd-Fe} = 0$ and $D_{Gd-Fe} < 0$, and in a ferromagnet. For a given DMI value, the radius of skyrmion is also different for the three scenarios, where smallest skyrmions are found with $D_{Gd-Fe} > 0$, and the largest skyrmions are found with $D_{Gd-Fe} < 0$.

To understand the intriguing behavior of skyrmion's size in ferrimagnet, in-plane spin configurations and the chirality of skyrmion's wall are investigated. **Fig. 3** summarizes the chirality of skyrmion wall at 0 K. With $D_{Gd-Gd}$, $D_{Fe-Fe} > 0$ and $D_{Gd-Fe} = 0$, for Fe sublattice, the spins in the skyrmion's wall are turning counter-clockwise. For Gd sublattice, the spins in the skyrmion's wall are also turning counter-clockwise. This can be explained by the dominance of exchange interaction in the system. Antiferromagnetic couplings between Gd and Fe align the spins of Gd and Fe in nearly antiparallel direction, with small canting due to presence of DMI. Identical behavior is observed with $D_{Gd-Gd}$, $D_{Fe-Fe} > 0$ and $D_{Gd-Fe} < 0$, where spins in both Gd and Fe sublattice are turning counter-clockwise across the skyrmion's wall. With $D_{Gd-Gd}$, $D_{Fe-Fe} > 0$ and $D_{Gd-Fe} > 0$, the chirality of skyrmion's wall is opposite to what observed in $D_{Gd-Fe} = 0$ and $D_{Gd-Fe} < 0$. The spins in both Gd and Fe sublattice are turning clockwise across the skyrmion's wall.

In order to determine the reason behind the change in chirality, the total DMI energies of each nearest neighbor pair are computed using equilibrium configurations at 0 K. **Table 2** summarizes the sign of total DMI energies of different nearest neighbor pair. With $D_{Gd-Gd}$, $D_{Fe-Fe} > 0$ and $D_{Gd-Fe} = 0$, the total DMI energy between Gd and Gd pair $E_{DMI}(Gd-Gd)$ and Fe and Fe pair $E_{DMI}(Fe-Fe)$ is negative, and the total DMI energy between Gd and Fe pair $E_{DMI}(Gd-Fe)$ is zero. This means that with $D_{Gd-Gd}$, $D_{Fe-Fe} > 0$, it is energetically favorable for spins to turn counterclockwise across skyrmion's wall. $E_{DMI}(Gd-Fe)$ is zero because $D_{Gd-Fe}$ is set to zero. With $D_{Gd-Gd}$, $D_{Fe-Fe} > 0$ and $D_{Gd-Fe} > 0$, $E_{DMI}(Gd-Gd)$ and $E_{DMI}(Fe-Fe)$ is positive, while $E_{DMI}(Gd-Fe)$ is negative. This implies that it is energetically favorable for Gd-Fe pair to turn clockwise across skyrmion's wall, but it is energetically unfavorable for Gd-Gd and Fe-Fe pair to do so. This means that in a ferrimagnet, if the DMI of ferromagnetic pair and antiferromagnetic pair has the same sign, cancellation of DMI occurs because it is preferable for ferromagnetic pair to turn in opposite direction of antiferromagnetic pair. With $D_{Gd-Gd}$, $D_{Fe-Fe} > 0$ and $D_{Gd-Fe} < 0$, all three terms $E_{DMI}$(Gd-Gd), $E_{DMI}$(Fe-Fe) and $E_{DMI}$(Gd-Fe) is negative, so turning counterclockwise is energy favorable for both ferromagnetic pair and antiferromagnetic pair in a ferrimagnet. These differences in sign of total DMI energy also explain the size of skyrmion in all three scenarios. For a given DMI, skyrmions are smallest for $D_{Gd-Fe} > 0$ because cancellation in DMI leads to reduction in DMI effectiveness in the sample. $D_{Gd-Fe} < 0$ scenario has the largest skyrmions because both ferromagnetic and antiferromagnetic are contributing to formation of a skyrmion, which means DMI is stronger overall. The trend of skyrmion's radius in $D_{Gd-Fe} > 0$ scenario can also be explained by cancellation of DMI between ferromagnetic and antiferromagnetic pair. As DMI becomes larger, more cancellation in DMI leads to smaller skyrmion. Thus, with large DMI, skyrmion decreases as DMI increases in the case of $D_{Gd-Fe} > 0$.

To determine the viability of using RE-TM alloys for skyrmion devices, simulations are also carried out at 300 K. Samples of 50.7 nm x 50.7 nm x 4.8 nm are simulated using atomistic stochastic LLG equation in Eq. (2). Since DMI is known decay away from the interface[52-55], an exponential decay DMI is employed in the simulation. **Fig. 4** shows the functional form of exponential decay DMI used in the simulation. In this model, DMI remains constant within 5 Å of the top and bottom interface, and start to decay exponential at 5 Å away from the interface.

**Fig. 5** summarizes the results of equilibrium spin configuration at 300 K. only ferrimagnetic states are observed with $D_{Gd-Fe} > 0$. As discussed earlier, with $D_{Gd-Fe} > 0$, cancellation of DMI between ferromagnetic and antiferromagnetic pair leads to unfavorable conditions for skyrmion formation. For the case of $D_{Gd-Fe} = 0$ and $D_{Gd-Fe} < 0$, small skyrmions of near 10 nm are found in the case of $D_{Gd-Fe} = 0$ and $D_{Gd-Fe} < 0$. Skyrmions this small are very promising for improving density and efficiency in skyrmion based devices. **Fig. 6** shows a comparison between atomistic simulation of GdFe and micromagnetic simulation of an equivalent ferromagnet. Using the same exponential decay DMI, much larger interfacial DMIis required to obtain skyrmion in the micromagnetic simulation of an equivalent ferromagnet. This demonstrates that internal spin structure in a ferrimagnet is essential to prolong the effect of DMI away from the interface. This DMI robustness in a ferrimagnet can be explained by the antiferromagnetic coupling between the two sublattices. Even without the presence of DMI, the spins in the Gd sublattice are known to be canted at room temperature[31]. With the presence of DMI, the spins in the Gd sublattice are easily guided by DMI and leads to formation of skyrmions. Thus, antiferromagnetic couplings can help to extend the influence of DMI, and increase stability of small skyrmions at room temperature.

**Conclusions**

Effect of interfacial DMI is investigated in amorphous ferrimagnetic GdFe using atomistic stochastic LLG algorithm. Three scenarios for the sign of DMI between Gd and Fe are considered. It is revealed that for a ferrimagnet, if the DMI between ferromagnetic pair and antiferromagnetic pair has the same sign, it leads to cancellation in DMI, and it is unfavorable for skyrmion formations. If the DMI between ferromagnetic pair and antiferromagnetic pair has opposite sign, it is advantageous for skyrmion formation, and small skyrmions of near ~10 nm are found to be stable at room temperature with exponential decay DMI. The antiferromagnetic couplings in ferrimagnet are uncovered to help extend the influence of DMI in thicker samples of ~ 5 nm. This discovery implies that antiferromagnetic coupling in ferrimagnet and antiferromagnet provides a favorable environment to stabilize small skyrmion at room temperature, which is an important recipe in developing high density and high efficiency skyrmion based device.

**Acknowledgements:**

This work is supported by DARPA

| Parameter | Value |
|---|---|
| **Gyromagnetic ratio (ϒ)** | 2.0023193 |
| Gilbert Damping (α) | 0.05 |
| Gd moment ($\mu_{Gd}$) | 7.63 $\mu_B$ |
| Fe moment ($\mu_{Fe}$) | 2.217 $\mu_B$ |
| Gd-Gd exchange constant ($J_{Gd-Gd}$) | 1.26 x $10^{-21}$ J |
| Fe-Fe exchange constant ($J_{Fe-Fe}$) | 3.82 x $10^{-21}$ J |
| Gd-Fe exchange constant ($J_{Gd-Fe}$) | -1.09 x $10^{-21}$ J |

**Table. 1 Values of parameters used in the simulation.**

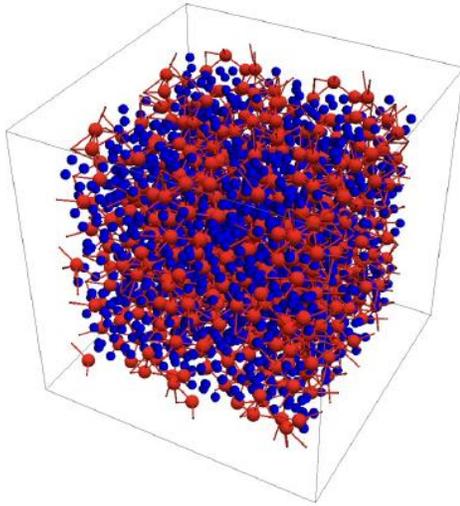

**Figure 1 Amorphous structure of $RE_{25}TM_{75}$ from ab initio molecular dynamics calculations.** Red atoms are rare-earth, and blue atoms are transitional-metal.

**Results and Discussion**

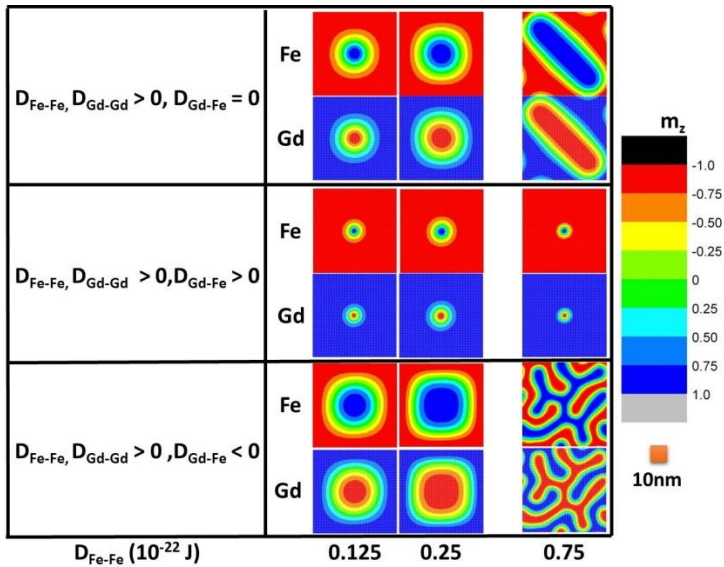

**Figure 2 Equilibrium spin configurations for various DMI (uniform DMI) at 0K for three scenarios of $D_{Gd-Fe}$.** Parameters used here are listed in Table 1 with Magnetic field $\mu_0 H$ is 0.01 T, anisotropy energy K is 0.3 x $10^5$ J/$m^3$ (distributed within a 45 degree cone), and simulation space

is 50.7 nm x 50.7 nm x 1.6 nm. For $D_{Fe-Fe}$ = 0.25 x $10^{-22}$ J, $D_{Gd-Gd}$ = 2.89 x $10^{-22}$ J, and $|D_{Gd-Fe}|$ = 0 or 0.85 x $10^{-22}$ J. For $D_{Gd-Fe}$ = 0 and $D_{Gd-Fe}$ < 0, the skyrmions size increases with DMI. On the other hand, for $D_{Gd-Fe}$ > 0, the skyrmion size first increases with DMI at small DMI, then decreases at larger DMI value.

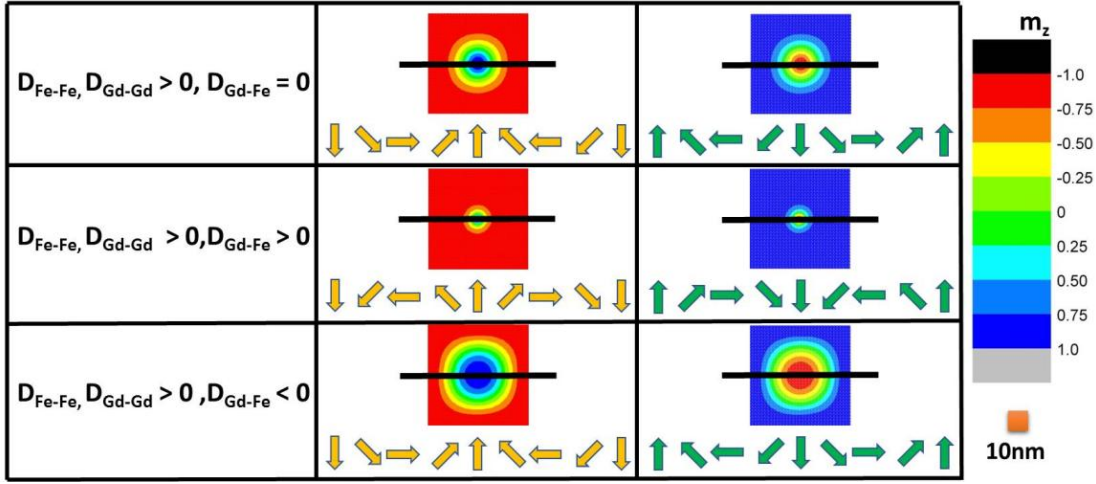

**Figure 3 Skyrmions wall chirality for three scenarios of $D_{Gd-Fe}$.** For $D_{Gd-Fe}$ = 0 and $D_{Gd-Fe}$ < 0, the skyrmions wall is rotating counter-clockwise. On the other hand, for $D_{Gd-Fe}$ > 0, the skyrmion wall is rotating clockwise.

| Scenario | $E_{DMI}$(Gd-Gd) | $E_{DMI}$(Fe-Fe) | $E_{DMI}$(Gd-Fe) |
|---|---|---|---|
| $D_{Gd-Gd}$, $D_{Fe-Fe}$ > 0, $D_{Gd-Fe}$ = 0 | - | - | 0 |
| $D_{Gd-Gd}$, $D_{Fe-Fe}$ > 0, $D_{Gd-Fe}$ > 0 | + | + | - |
| $D_{Gd-Gd}$, $D_{Fe-Fe}$ > 0, $D_{Gd-Fe}$ < 0 | - | - | - |

**Table. 2 Sign of total DMI energy $E_{DMI}$ computed from equilibrium spin configurations at 0 K.**

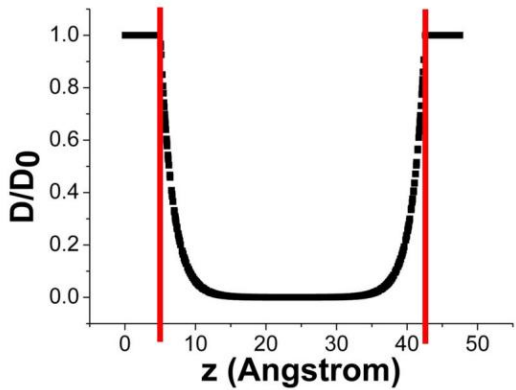

**Figure 4 Plot of exponential decay DMI as function of distance from bottom interface (z).** In this model, DMI remains constant within 5 Å of the top and bottom interface, as indicated by the red line. At the center of the 4.8 nm sample, the strength of DMI decays exponentially as shown.

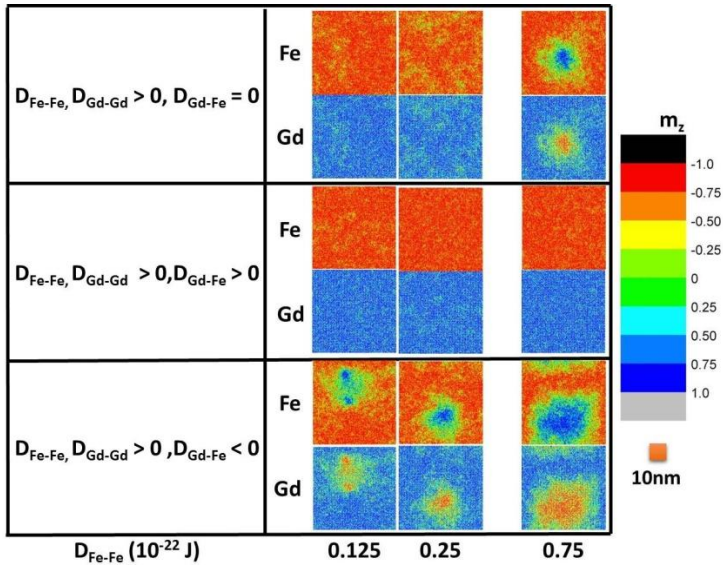

**Figure 5 Equilibrium spin configurations for various DMI (exponential decay DMI) at 300K for three scenarios of $D_{Gd-Fe}$.** Parameters used here are listed in Table 1 with Magnetic field $\mu_0 H$ is 0.01 T, anisotropy energy K is $0.3 \times 10^5$ J/m$^3$ (distributed within a 45 degree cone), and simulation space is 50.7 nm x 50.7 nm x 4.8 nm. For $D_{Fe-Fe} = 0.25 \times 10^{-22}$ J, $D_{Gd-Gd} = 2.89 \times 10^{-22}$ J, and $|D_{Gd-Fe}| = 0$ or $0.85 \times 10^{-22}$ J. At 300K with exponential decay DMI, skyrmions are only found to exist with $D_{Gd-Fe} = 0$ and $D_{Gd-Fe} < 0$.

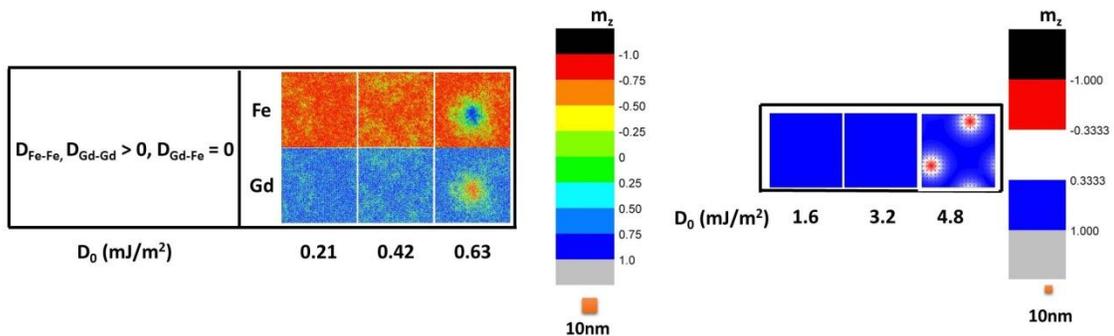

**Figure 6 Equilibrium spin configurations from atomistic simulation of GdFe at 300 K (left) and micromagnetic simulation of an equivalent ferromagnet at 0 K (right).**